# Investigation of the effects of transient heat loads on plasma-facing materials in Tokamaks

Ali Masoudi [a], Davoud Iraji [a,*], Chapar Rasouli [b]

[a] *Department of Physics and Energy Engineering, Amirkabir University of Technology, Hafez Ave, Valiasr Square, Tehran, 1591634311, Iran*

[b] *Plasma Physics and Nuclear Fusion Research School, Nuclear Science and Technology Research Institute, North Kargar Ave, Tehran, 14155-1339, Iran*

**Keywords**

Plasma-facing materials
Plasma-facing components
Edge localized modes
Tokamak
Plasma

**Abstract**

Nuclear fusion devices are constantly under the threat of malfunctions coming from the damages of plasma-facing materials due to being affected by thermal heat loads. The frequent heat loads during some transient events in large-scale Tokamaks have always been a great concern for researchers. In ITER, the heat load of $GW/m^2$ is estimated to impose plasma-facing components during edge localized modes, beside the Tokamak steady state load which is about 20 $MW/m^2$. Moreover, there are also other transient thermal loads occurring due to off-normal operation of ITER such as vertical displacement events or disruptions, at the orders of hundreds of $MW/m^2$ and tens of $GW/m^2$, respectively. These loads are great enough to result in severe damages of plasma-facing materials. In this study, the facture of tungsten material under the heat loads of Tokamaks is simulated and the results are presented.

## Graphical Abstract

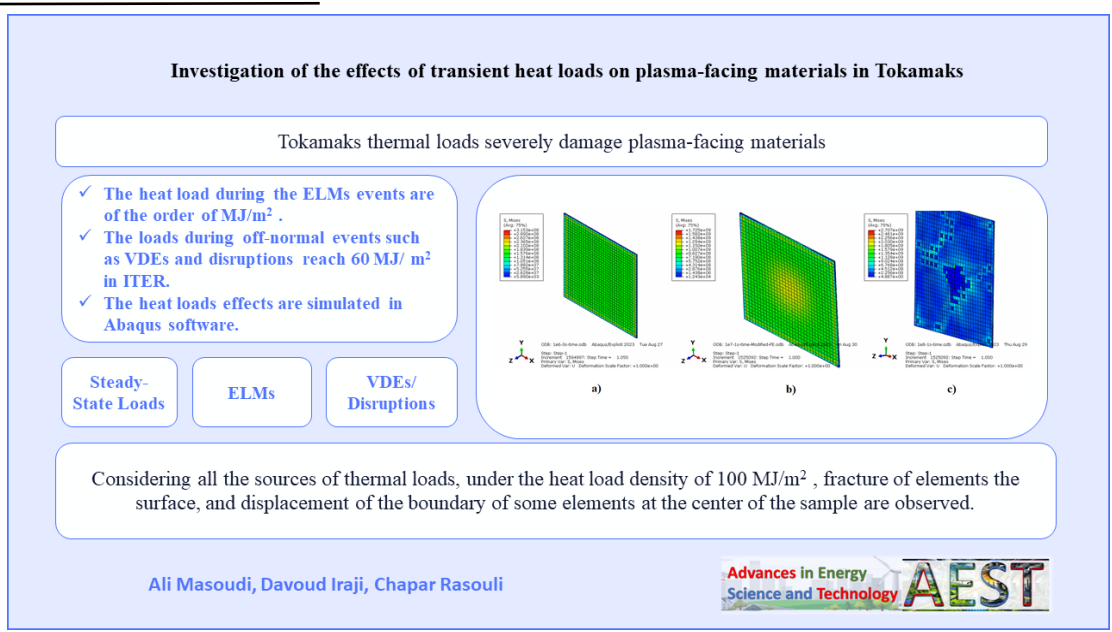

# 1. Introduction

The ever-increasing need of humankind for abundant energy sources, while preserving the environment, has convinced humanity to invest in, research, and develop nuclear fusion energy, akin to the energy produced by processes occurring in the sun. Indeed, significant progress has been made in this field in recent years and months [1-7].

PFCs (Plasma-Facing Components) are constantly in direct contact with heat loads from particles present at the edge of the plasma. In addition to these steady-state loads, which impose a background thermal load on the PFCs at the average density and temperature of the plasma, fireballs are also ejected from the core plasma during events known as ELMs (Edge Localized Modes) [8].

The steady-state and transient heat loads on Plasma-Facing Material (PFMs) lead to the material's erosion. As a result, the eroded material can become ionized and enter the plasma, causing contamination and subsequent cooling of the plasma through Bremsstrahlung radiation. Generally, plasma-wall interactions (PWI) have a significant impact on many operational aspects of a fusion reactor, including plasma performance, the lifespan of the wall material, reactor efficiency, fuel (tritium) retention, and consequently, safety and operational considerations [9].

In summary, the thermal load generated by plasma, wherever it comes into contact with material, can alter the material's structure. The consequences of such structural changes have the potential to disrupt the functionality of a system [10]. However, this disruption becomes more pronounced on larger scales and can lead to significant damage that may be difficult or even impossible to repair.

Therefore, the appeal and necessity of the proposed topic become evident here, as research in this field is continuously evolving and requires newer investigations and updated data [11]. On the other hand, since conducting experiments to assess the effects of thermal loads in large-scale devices such as ITER or DEMO is not feasible before their operational launch, and after their activation, there will be no opportunity for further examination or the possibility of replacing or repairing components, it is essential to conduct enough research using available tools. By extrapolating the results to operational and large-scale devices, multiple suitable options can be proposed in this regard.

It was mentioned that the loads during the ELMs affect the PFMs with nearly the same density and temperature as the core plasma, significantly adding to the steady-state thermal load. Therefore, a PFC designed to withstand the ELMs will inherently be compatible with the thermal load from the edge plasma. This underscores the importance of investigating the effects of ELMs on the PFCs.

*Hassanein* et al. have presented a paper that discusses the internal design of ITER during transient events such as ELMs under normal operation and disruptions under off-normal operation, as well as the significant challenges related to the design of the PFCs [12].

In some reactor designs with fully metallic first walls, the presence of dust can be reduced. However, when the first wall is entirely made of tungsten, its exposure to thermal loads can lead to significant concerns due to its high atomic number. The potential effects of tungsten dust on the device's performance are notable, and in future reactors, the amount of dust in the chamber must comply with nuclear licensing requirements. A summary of historical research on dust in fusion devices with carbon-based PFCs, the formation of metallic dust, its movement, and its interaction with fusion plasma and PFCs has been presented in a paper by *Ratynskaia* et al. [13].

This study aims to investigate the ELMs events in Tokamaks and simulate the effect of Tokamak thermal loads on the PFMs and the paper is structured as follows: The theoretical basis and study of the behavior of the ELMs is described in Section 2.

Section 3 demonstrates the simulation method of transient heat loads on the PFMs. Eventually, the conclusion of the study is presented in section 4.

## 2. Theoretical basis

When a tokamak operates in L-mode (Low-confinement Mode), the small-scale turbulence present at the edge of the plasma causes the loss of particles and energy, thereby reducing the pressure of the core plasma. By increasing the heating power and transitioning to H-mode (High-confinement Mode), the turbulence at the plasma edge is significantly reduced. This mechanism is accompanied by a large-scale suppression of the edge plasma current, which decreases the transport of particles and energy from the core plasma and effectively creates an edge transport barrier. This barrier leads to a steep pressure gradient at the plasma edge. This region acts as an insulating layer around the plasma, increasing the overall core pressure and, consequently, improving confinement. However, despite the benefits of edge transport barriers, if their pressure gradient exceeds a certain threshold, they can trigger explosive instabilities known as ELMs [14].

Therefore, in enhanced confinement plasma (H-mode), the steep gradients in density and temperature at the plasma edge trigger ELMs, which are MHD (MagnetoHydroDynamic) instabilities characterized by rapid and repetitive plasma eruptions. These ELMs cause the expulsion of energy and particles from the core plasma, transferring them to the plasma edge. The large particle and heat fluxes resulting from these instabilities impose significant loads on reactor components and are predicted to cause severe erosion of the surrounding material surfaces. As a result, the ELMs are one of the major concerns for future tokamaks, such as ITER [15]. Actually, the primary challenge in designing future fusion reactors lies in their performance during off-normal events, such as disruptions and ELM events. A successful reactor design must endure numerous transient events without sustaining significant damage, such as melting or structural changes [12].

To prevent severe damage and ensure the successful operation of ITER, its divertor may require significant modifications or innovative design approaches. When a thermal load is applied to the divertor, a secondary plasma forms above its surface, extending along the magnetic field lines into the Scrape-Off Layer (SOL). This secondary plasma separates particles from disruptions or ELM events originating from the core plasma and directed toward the divertor. In the case of ELMs, it can penetrate the core plasma, effectively leading to a complete loss of confinement. The energy from particles in transient events is primarily converted into two significant secondary heat sources: photon radiation and fluxes of scattered particles. These can cause severe damage or erosion to many internal or hidden components that are not directly exposed to disruptions or ELMs [12].

A prominent characteristic of ELMs is the millisecond-scale spikes observed in the $H_\alpha$ emission signal from the plasma boundary. Type I ELMs, as shown in **Fig. 1**, which corresponds to a discharge in JET with neutral beam injection heating, are observed in nearly all devices [16]. According to the figure, it is evident that 47 Type I ELM events occurred in JET over a period of 6 seconds. During this discharge, with a power of approximately 13 $MW$, the occurrence of ELMs caused particles with a density of $8 \times 10^{19}$ $m^{-3}$ and energy of about $8 \times 10^5$ $J$ to collide with the PFCs.

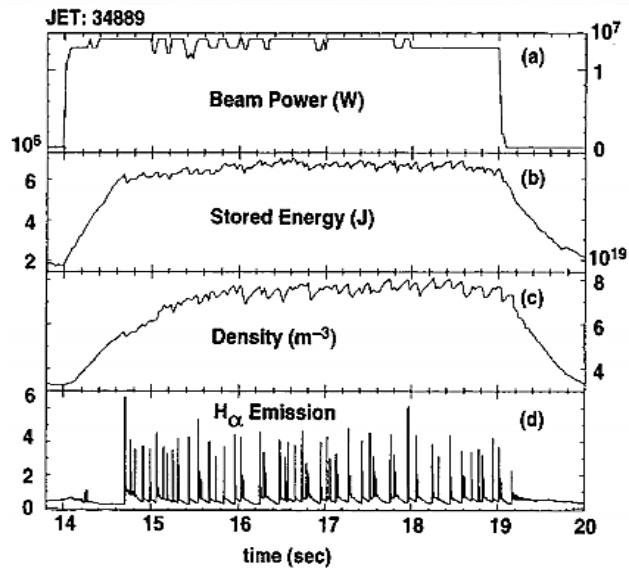

**Fig. 1.** An H-mode discharge in JET demonstrating Type I ELM events [16].

ELM events are classified based on their temporal (and energy) characteristics. The effective energy pulse applied by the ELMs in the plasma boundary region consists of a rise and fall phase. For example, in the HL-2A tokamak (**Fig. 2**), the rise time for the appearance of the ELM energy ranges between 100 $\mu s$ and 400 $\mu s$, while the fall time is 1.5 to 4 times longer than the corresponding rise time [17].

In **Fig. 3a**, the fall time of ELMs is plotted against the rise time. In **Fig. 3b**, the vertical axis represents the ratio of ELM energy during the fall phase to its energy during the rise phase, while the horizontal axis shows the rise time. This energy ratio varies from 2 to 5, indicating that the ELM energy during the fall phase is greater than during the rise phase [17].

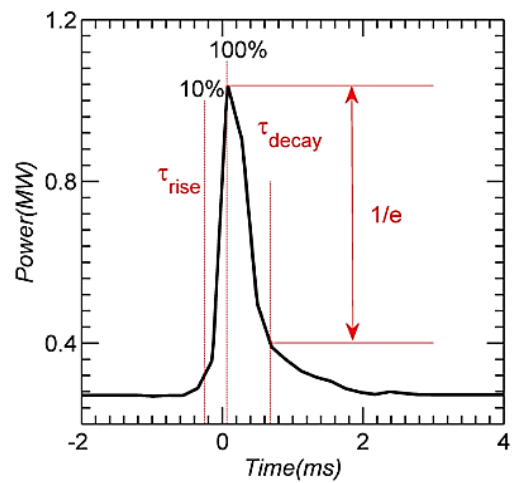

**Fig. 2.** The thermal evolution of the average power transfer across 7 consecutive ELM events [17].

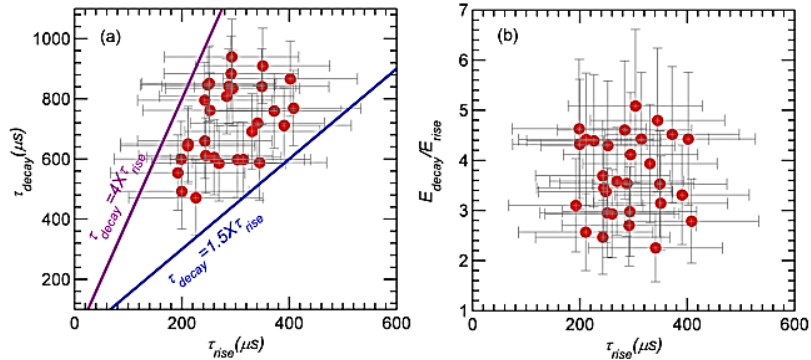

**Fig. 3.** Temporal characteristics of Type I ELMs in HL-2A [17].

Regarding this transient event, it can be said that an ELM is a relatively fast MHD event (100–200 $\mu s$) described by electrostatic and magnetic turbulence. **Fig. 4** shows the electron density and temperature profiles before and after an ELM event, and it is evident that during the ELM, the electron density at the separatrix ($\delta R = 0$) drops sharply [16].

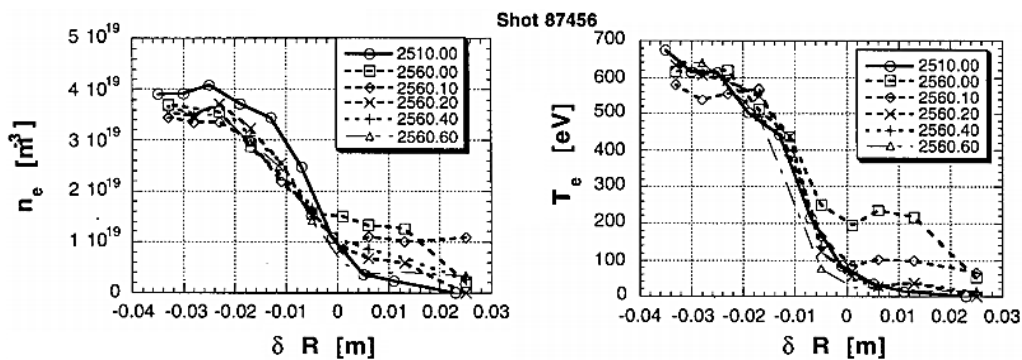

**Fig. 4.** Radial profiles of electron temperature and density before (2510 $ms$) and during (2559.9 $ms$) a Type I ELM in DIII-D [16].

One of the primary ways to identify Type I ELMs is that their frequency increases with heating power. Additionally, it has been demonstrated in DIII-D and JET that the ELM frequency decreases significantly with plasma current (**Fig. 5**) [16].

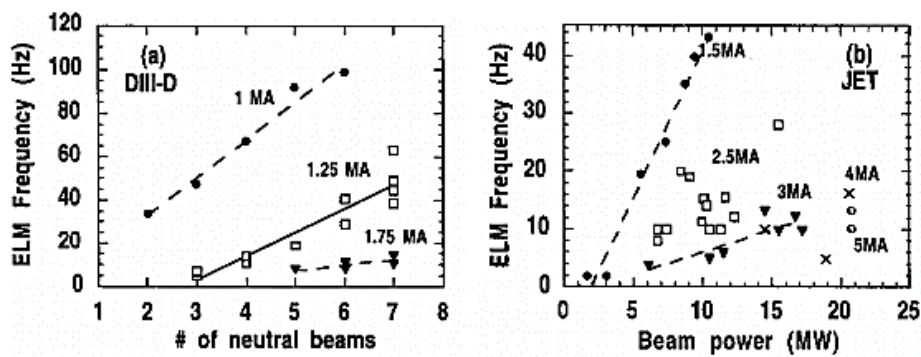

**Fig. 5.** Dependence of Type I ELM frequency on plasma current and beam power in DIII-D and JET [16].

Parameters such as electron thermal conduction time (100 $\mu s$ or less), ion current flow time (~1 $ms$), and neutral gas transit time to the divertor (~2–10 $ms$) are among the temporal characteristics of the ELMs. The energy flow is primarily propagated by electron conduction along the magnetic field lines. Therefore, the thermal wave caused by electrons first reaches the wall or divertor surface, and it takes time for energy to transfer from the separatrix. The ion current entering the divertor from the main plasma is limited by the ion sound speed and is thus slower, while the initial thermal pulse can increase the ionization of neutral particles. Finally, after ions are transported to the divertor target, particles begin to recycle, and a new equilibrium is established within a few milliseconds [16].

The divertor heat pulse resulting from Type I ELMs in ASDEX-U is also presented in **Fig. 6**. The heat pulses in ASDEX-U are generally smaller than those in DIII-D and also smaller than those in JT-60U [16].

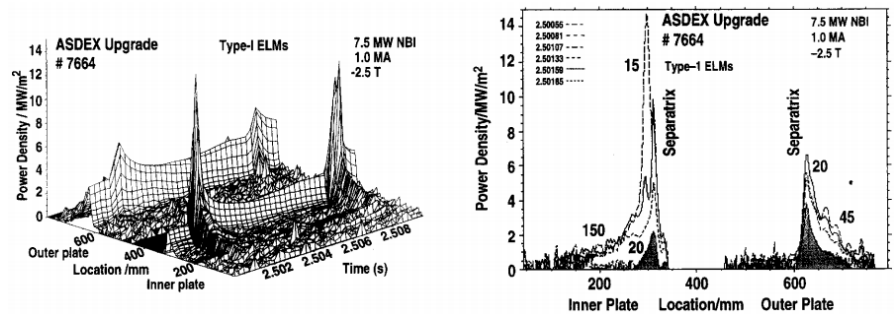

Fig. 6. Divertor heat flux profiles during ELMs in AUG [16].

Since, during an ELM event, a portion of the core plasma is ejected, the thermal load and conditions of these events at the time of wall impact are nearly identical to the parameters of the core plasma. Therefore, the temperature and density profiles of the core plasma can be referenced to extract the necessary data. The radial distribution of these parameters also provides essential information about the edge plasma. In **Fig. 7**, the $D^+$ flux (solid blue), D flux (black dashed), flux of large tungsten ions (solid red line), and net ion flux (green plus signs) in WEST with ICRH heating at 6 $MW$ are shown. The temperature and density profiles of the plasma in TFTR are also presented in **Fig. 8** [18], [19].

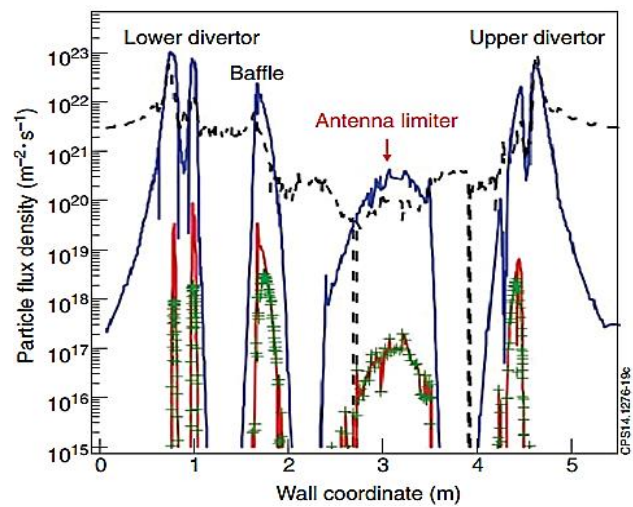

Fig. 7. Particle flux density as a function of poloidal position in WEST Tokamak [18].

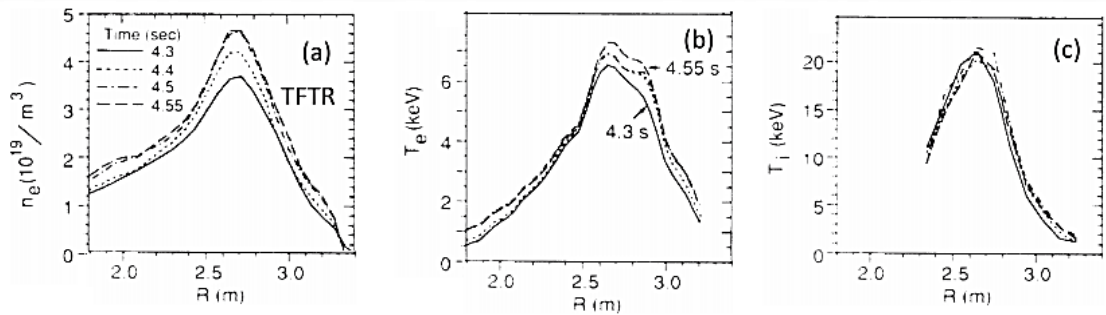

**Fig. 8.** Radial profiles of electron density, electron temperature, and ion temperature in TFTR [19].

The kinetic profiles of the core plasma for the JET tokamak (temporal evolution of electron temperature at the plasma center and electron plasma energy) are also presented in **Fig. 9**. These profiles remain unchanged with gas injection from the top, divertor, or OMP[1]. In this figure, the auxiliary heating power, $D_2$ gas injection rate, electron temperature at the plasma center, total electron energy, as well as electron density and temperature profiles are shown, respectively [20].

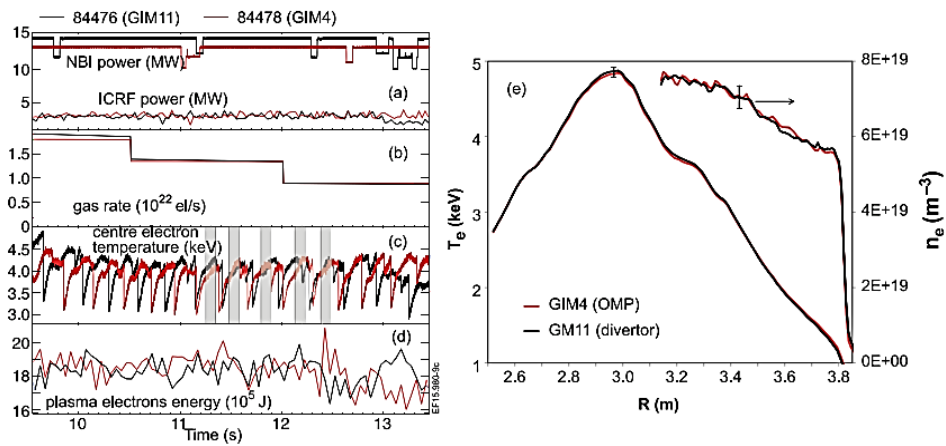

**Fig. 9.** Temporal evolution of the electron kinetic profile in JET [20].

The background plasma parameters in the JET tokamak at a net input power of 14 *MW* are also presented in **Fig. 10**. The average density at the separatrix is $2\times10^{19}$ *m⁻³*. In this figure, the vertical plasma transport coefficients ($D_\perp$ and $\chi_\perp$), comparisons of midplane electron density (error bars correspond to reflectometer data), comparisons of midplane electron temperature, and comparisons of particle flux entering the divertor (around the outer strike point at the Roof Baffle) are shown, respectively [20].

---

[1] Outer Mid-Plane

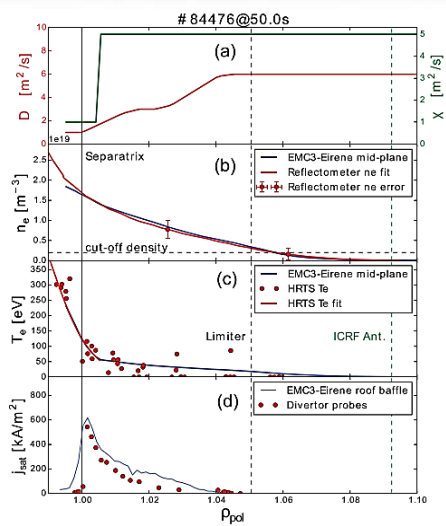

**Fig. 10.** Matching plasma parameters with experimental data in JET [20].

Another parameter of tokamak plasma is its pressure. The operating pressure of fusion reactors in the burn phase is on the order of $2\times10^{-2}$ *Pa*, and in the stationary phase, it is 0.1 *Pa* [21]. The plasma pressure in ITER can be extracted from the data in **Fig. 11**, which is obtained from simulations of ITER parameters at three fusion power levels: 375 MW, 400 *MW*, and 415 MW [22]. The plasma pressure in DIII-D is also plotted in **Fig. 12**.

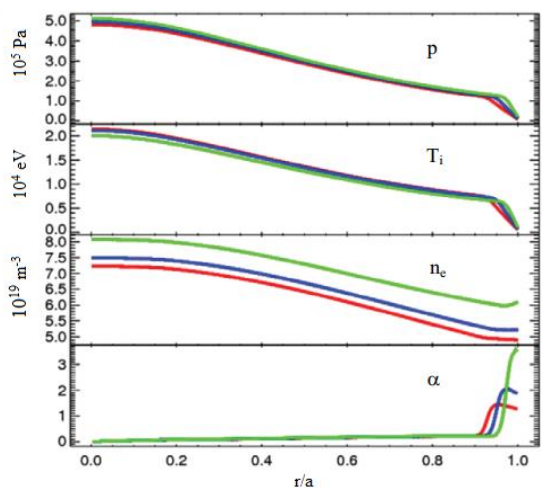

**Fig. 11.** Profiles of pressure, ion temperature, electron density, and normalized pressure gradient in ITER [22].

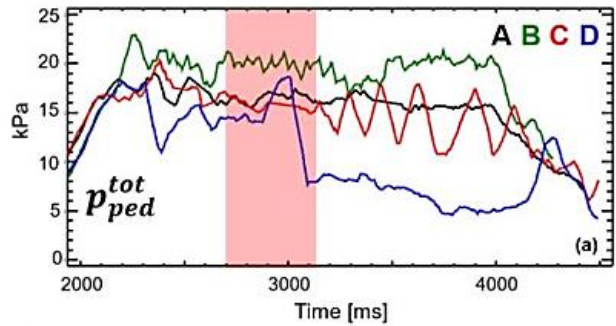

**Fig. 12.** Temporal evolution of base pressure for radiative powers of 0, 4.5 MW, 7.5 MW, and 8.5 MW (A to D) for the lower divertor in DIII-D [23].

According to **Fig. 13** the neutral gas pressure in EAST is approximately 3.5 *Pa*. In this tokamak, the average electron density in the core plasma reaches $1.3\times10^{19}$ *m$^{-3}$*, and the electron temperature in the core plasma is about 3 *keV*, although these values are significantly lower in the edge plasma [24].

In general, in the fusion process, the temperature of the core plasma reaches around 1 *keV*, but it drops sharply in the SOL. However, this temperature is still high enough to melt PFC materials. The rotation speed of dust particles in low-temperature plasma can reach up to $10^5$ *rad/s*, and in high-temperature plasma, this angular velocity can increase to $10^9$ *rad/s* due to the Lorentz force [24].

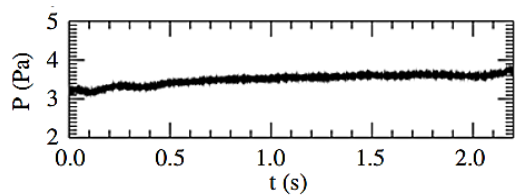

**Fig. 13.** Neutral gas pressure in EAST [24].

In summary, considering the discharge duration in each fusion reactor, the number of ELM events during this time, and the energy deposited on PFC components in each event, it is clear that without the use of suitable materials, the components will lose their functionality over time and disrupt the reactor's operation. As of the preparation of this document, the EAST tokamak holds the world record for plasma confinement duration, having maintained a steady plasma pulse for 1053 seconds. In this tokamak, the peak steady-state heat flux to the divertor target plates was approximately 3 *MW/m²*. Prior to this, discharges lasting up to 6 minutes have been achieved in EAST and other well-known tokamaks such as WEST and HT-7 [25]. In the HL-2A tokamak, the peak heat flux to the divertor during an ELM event is approximately 1.5–3.2 *MW/m²*, and the plasma contact area was about 0.5–0.7 *m²* [17].

During a Type I ELM event, which is the most severe type of ELM, up to 10% of the core plasma energy is transferred to the surface of the PFCs. Additionally, various studies have consistently estimated the total energy from disruptions in ITER to be in the range of 120–175 *MJ*, with approximately 10% of this value, around 17.5 *MJ*, predicted for the ELMs. This represents an extremely high thermal load. If the ELM repetition rate is assumed to be even 1–10 *Hz*, it can cause severe heating of the PFCs, leading to surface cracking or melting. Beyond damaging the components, the resulting sputtering of particles contaminates the primary plasma. Furthermore, despite efforts to mitigate ELM thermal loads, recent estimates suggest that even mitigated ELM events can impose energy densities of up to 1 *MJ/m²* (at a frequency of 50 *Hz*) on the ITER PFCs [26-27].

In other words, the standard plasma discharge pulse duration in ITER will be 450 seconds, during which the PFCs will be exposed to quasi-steady-state heat loads of up to 20 *MW/m²* (causing recrystallization and joint failure) along with rapid and extremely intense transient heat loads. The unmitigated type of these loads can deliver energy on the order of tens of *GW/m²* during ELM events lasting 0.5 *ms*, with frequencies of several *Hz* (leading to cracking, melting of components, and dust formation). In some studies, the duration of these events for the current ITER design is reported as 1 *ms* [28]. Both the quasi-steady-state heat loads of plasma discharges and the short thermal spikes from ELMs are normal operational scenarios that simultaneously apply their energy to the divertor and PFCs. However, ITER is a device that must also investigate new plasma regimes, and events such as plasma disruptions—sudden termination of magnetically confined plasma—or vertical displacement events (VDEs)—upward or downward movement of hot plasma due to improper plasma positioning system operation—may also occur. Therefore, careful selection of materials suitable for long-term plasma operation in ITER is essential. The density and duration of pulses expected from these events depend on the tokamak's dimensions, and the wall load during disruptions can reach 30 *MJ/m²* for 2–5 *ms*, while for VDEs, it can be 60 *MJ/m²* for 100–300 *ms* in ITER. **Fig. 14** shows the power density versus event duration in ITER. The turquoise region represents operational conditions during abnormal events, where material quality degrades irreversibly. This irreversible damage depends on the number of pulses (for ELMs) [29].

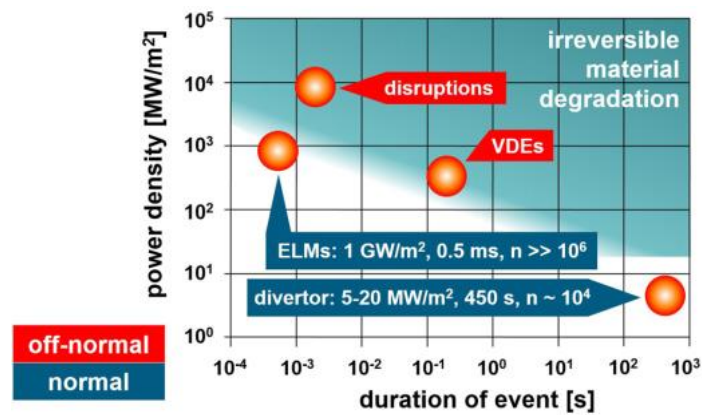

**Fig. 14.** Heat loads on the divertor target in ITER [29].

The effect of damage on actively cooled divertor components for ITER is shown in **Fig. 15**. The left side of the figure illustrates relatively severe macroscopic damage caused by cyclic heat loads with 1000 cycles at an absorbed power density of 20 MW/m². The right side shows severe damage to the plasma-facing armor due to transient heat loads with a very high number of cycles, where $10^5$ pulses with a heat flux factor of 12 $MW \cdot m^{-2} \cdot s^{0.5}$ were applied to a tungsten test sample initially heated to 700 °C using intense electron beam pulses. This sample exhibits the formation of a dense network of cracks on the plasma-exposed surface. The typical ELM frequency in ITER is 25 *Hz*, and its flat-top phase lasts 400 seconds, resulting in $10^4$ ELM events per discharge. Therefore, during operation with 10 standard plasma discharges in ITER, $10^5$ ELM events occur [28]. Thus, sufficient studies must be conducted on the performance of PFC materials under this number of ELM events and the significant energy they generate.

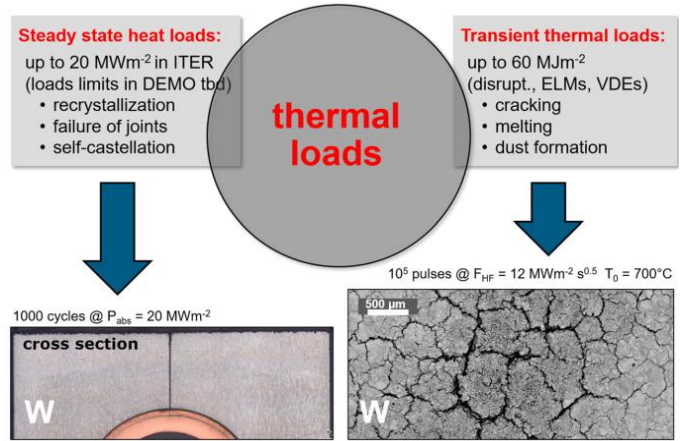

**Fig. 15.** Damage at the divertor of ITER caused by 1000 cycles of 20 *MW/m²* steady-state loads (the left side) and $10^5$ pulses of transient ELM event loads with a heat flux of 12 *MW. m⁻².s⁰·⁵* (the right side) [29].

## 3. Results and Discussion

As it was stated in previous sections, steady and transient thermal loads in Tokamaks can altogether result in significant damages to the PFCs, such as surface delamination, blister formation, or any other adverse effects.

In order to simulate heat load effects on PFMs, ABAQUS software was utilized in this study. In the first step, it is necessary to know the constitutive law for materials and find model inputs. The purpose is to simulate the crack formation on the surface of the PFM.

There are different behavior models for materials, and a famous damage model, which has been chosen, is *Johnson Cook* (JC) model. This model describes the visco-thermo-plastic behavior of materials and is usually applied in finite element studies. This law is shown through **Eq. (1)**, in which σ is the flow stress, *ε* is the equivalent plastic strain, $\dot{\varepsilon}$ strain rate, $\dot{\varepsilon}_0$ is reference strain rate, $T_r$ and $T_m$ are ambient reference temperature and melting temperature, respectively [29].

$$\sigma = \left[ A + B(\varepsilon)^n \right]\left[ 1 + C\ln(\frac{\dot{\varepsilon}}{\dot{\varepsilon}_0}) \right]\left[ 1 - \left( \frac{T - T_r}{T_m - T_r} \right)^m \right]. \quad (1)$$

**Eq. (2)** is used to define the initial failure strain [30].

$$\varepsilon_{oi}^{-pl} = \left[ d_1 + d_2 \exp(d_3 \frac{p}{q}) \right]\left[ 1 + d_4 \ln(\frac{\dot{\varepsilon}^{pl}}{\dot{\varepsilon}_0}) \right]\left[ 1 - d_5(T^*) \right]. \quad (2)$$

Where $\varepsilon_{oi}^{-pl}$ is equivalent plastic strain at the onset of damage, *p/q* is a dimensionless pressure-deviatoric stress ratio (*p* and *q* are pressure stress and von Mises stress, respectively), $T^*$ is non-dimensional temperature, and *(T − $T_r$)/($T_m$ − $T_r$)* and $d_i$ are the failure parameters [29].

For tungsten as a PFM with high specific strength and corrosion resistance, the constitutive law was extracted from references [30-32].

The considered sample is a homogeneous cube with a size of 1 *m* × 1 *m* × 0.01 *m* which a uniform heat load is applied from the front face. The temperature of 50 °C is applied as a boundary condition to the bottom surface. The surrounding faces are constrained from rotation and displacement. The time period of the simulation is 1 *s*.

This simulation gives Von Mises Stress distribution, which if it exceeds a criterion for each material, the material fracture is occurred. This criterion is of the order of *GPa* for usual PFMs such as tungsten.

As it is clear from **Fig. 16a**, the application of 1 MJ heat load to the sample leads to the maximum stress of about 300 *MPa*. By increasing the load to 10 *MJ*, the maximum Mises stress reaches 1.7 *GPa*, that doesn't result in the material failure yet (**Fig. 16b**). After applying 100 *MJ* to the sample surface, the maximum stress is 2.7 *GPa*, that causes the material damage which is appeared as deviation and distortion of the surface elements. This damage is an obvious sign of cracking and material delamination as demonstrated in **Fig. 16c**.

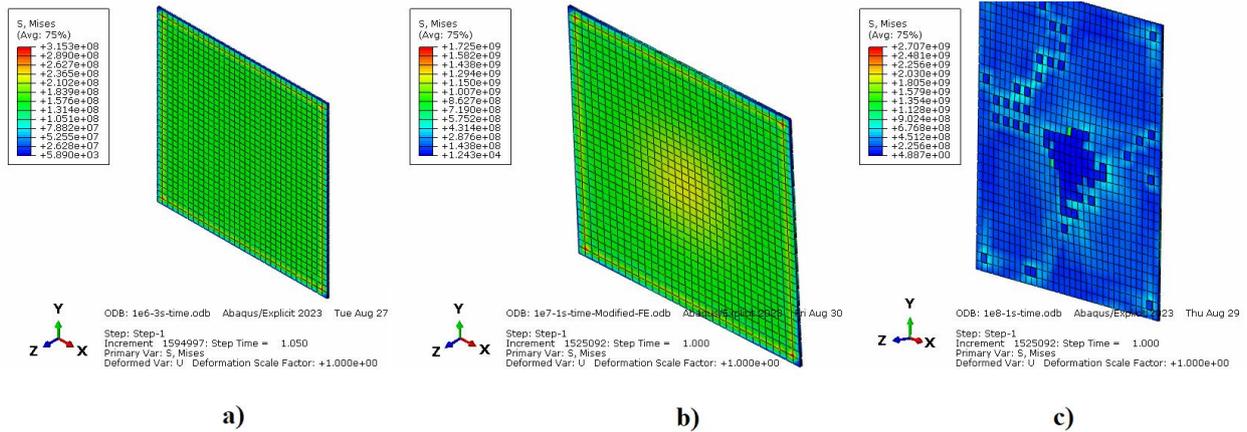

**Fig. 16.** Distribution of Mises stress on W surface after applying (a) 1 *MJ*, (b) 10 *MJ*, and (c) 100 *MJ* heat loads.

**Fig. 17** shows the magnified view of the sample surface under the heat load of 100 *MJ*. This figure demonstrates the fracture of elements on the first and second layers of the surface. Displacement of the boundary of some elements at the center of the sample can be due to swelling of these elements before the fracture and delamination.

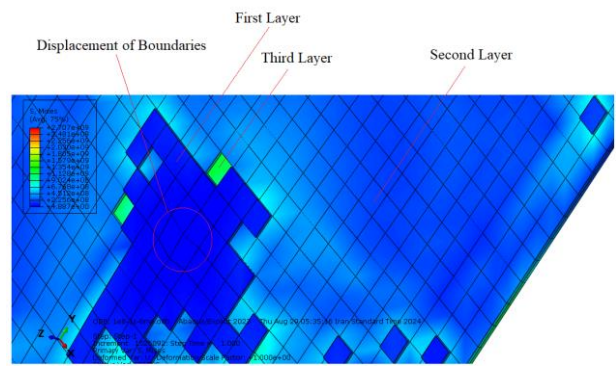

**Fig. 17.** Magnified view of the surface when the applied heat load is 100 *MJ*.

## 4. Conclusion

Plasma-facing materials in confinement fusion reactors are exposed to steady-state and transient thermal loads during normal and off-normal operations of the devices. These heat loads, besides the interaction of plasma particles with the materials and neutron bombardments, have made the study and design of the PFCs a challenging field of research. The large-scale devices such as ITER are not yet available to be used in order to conduct the research, and after the starting of their operation starts, there will be no time to try and error to repair or replace the damaged components or change the designs. Therefore, a logical way is to carry on every possible investigation, making use of each available device to consider all aspects of present designs and predict every possible issue.

In this study, the effect of thermal loads on the PFMs has been modeled and simulated using ABAQUS software. The Johnson-Cook method has been utilized to model the visco-thermo-plastic behavior of tungsten as the potential and most used PFM. Considering that the ELMs are the most challenging events in high confinement devices, causing irreparable damage to the PFCs, these events have been studied, and the order of their energy has been extracted to be used in the modeling of transient loads. Then, the transient heat loads near the orders of the energy release during the ELMs and some other off-normally occurring events, such as VDEs, have been applied to the modeled sample.

The simulation results show that the heat loads during these events can potentially damage the PFCs, and the damage appears as fractures of the elements, meaning the cracks on the surfaces and delamination of the tungsten sample.

## Author contributions

**A.M.** conceived the idea, conceptualized the research, performed the computer simulations, and wrote the paper. **D.I.** supervised the project, directed the research, and revised the manuscript. **C.R.** validated, investigated, administered the project, reviewed, and edited the manuscript.

## Declaration of competing interest

The authors declare that they have no known competing financial interests or personal relationships that could have appeared to influence the work reported in this paper.

## Data availability

Data will be made available on request.

## References

[1] G. Binfu, Rui Ding, H. Xie, L. Zeng, L. Zhang, B. Wang, C. Li, D. Zhu, R. Yan, and J. Chen, Plasma-facing components damage and its effects on plasma performance in EAST tokamak, Fusion Eng. Des. 156 (2020) 111616.

[2] Y. Corre, M. H. Aumeunier, A. Durif, J. Gaspar, K. Krieger, T. Loewenhoff, M. Richou, S. Ratynskaia, Q. Tichit, Y. Anquetin, R. Dejarnac, Testing of ITER-grade plasma facing units in the WEST tokamak: Progress in understanding heat loading and damage mechanisms, Nucl. Mater. Energy 37 (2023) 101546.

[3] I. V. Mazul, R. N. Giniyatulin, A. A. Kavin, N. V. Litunovskii, A. N. Makhankov, P. Yu Piskarev, V. N. Tanchuk, Plasma-Facing Components of the TRT Tokamak, Plasma Phys. Rep. 47 (2021) 1220–1237.

[4] A. Hakola, J. Likonen, A. Lahtinen, T. Vuoriheimo, M. Groth, H. Kumpulainen, M. Balden, K. Krieger, M. Mayer, T. Schwarz-Selinger, S. Brezinsek, Gross and net erosion balance of plasma-facing materials in full-W tokamaks, Nucl. Fusion 61 (2021) 116006.

[5] E. Tsitrone, B. Pegourie, J. P. Gunn, E. Bernard, V. Bruno, Y. Corre, L. Delpech, M. Diez, D. Douai, A. Ekedahl, N. Fedorczak, Investigation of plasma wall interactions between tungsten plasma facing components and helium plasmas in the WEST tokamak, Nucl. Fusion 62 (2022) 076028.

[6] S. D. Fedorovich, A. V. Karpov, V. P. Budaev, S. A. Grashin, M. K. Gubkin, M. V. Lukashevsky, I. V. Voinkova, K. A. Rogozin, A. A. Konkov, V. T. Quang, G. B. Vasiliev, High-heat flux tests of graphite in-vessel components of T-15MD tokamak with plasma and e-beam load, J. Phys.: Conf. Ser. 1683 (2020) 032012.

[7] Y. Lei, W. Fu, D. Zhu, X. Yang, C. Li, R. Ding, J. Chen, B. Yu, C. Xuan, J. Li, Damages of TZM as plasma facing material under transient heat load, Fusion Eng. Des. 194 (2023) 113864.

[8] Z. Guo, B. Yu, D. Zhu, C. Xuan, B. Gao, C. Li, R. Ding, J. Chen, C. Du, Analysis of the transient heat flux on divertor surface during plasma disruption in EAST, Nucl. Mater. Energy 36 (2023) 101503.

[9] J. Schmitz, Development of tungsten alloy plasma-facing materials for the fusion power plant, (PhD thesis), Universiteit Gent & Ruhr-Universität Bochum (Promotor: K. Crombé) (2020).

[10] Y. Ueda, J. W. Coenen, G. De Temmerman, R. P. Doerner, J. Linke, V. Philipps, E. Tsitrone, Research status and issues of tungsten plasma facing materials for ITER and beyond, Fusion Eng. Des. 89 (2014) 901–906.

[11] A. Masoudi, D. Iraji, Investigation of glow discharge plasma energy distribution using a gridded energy analyzer considering plasma-facing materials related processes, Fusion Eng. Des. 212 (2025) 114862.

[12] A. Hassanein, V. Sizyuk, Potential design problems for ITER fusion device, Sci. Rep. 11 (2021) 2069.

[13] S. Ratynskaia, A. Bortolon, S. I. Krasheninnikov, Dust and powder in fusion plasmas: recent developments in theory, modeling, and experiments, Rev. Mod. Plasma Phys. 6 (2022) 20.

[14] C. Ham, A. Kirk, S. Pamela, H. Wilson, Filamentary plasma eruptions and their control on the route to fusion energy, Nat. Rev. Phys. 2 (2020) 159–167.

[15] H. Zohm, Edge localized modes (ELMs), Plasma Phys. Control. Fusion 38 (1996) 105.

[16] D. N. Hill, A review of ELMs in divertor tokamaks, J. Nucl. Mater. 241–243 (1997) 182–198.

[17] J. M. Gao, L. Cai, X. L. Zou, W. Y. Zhong, T. Xu, J. Qin, Z. Zhang, Y. Liu, Type-I ELM power loads on the closed outer divertor targets in the HL-2A tokamak, Nucl. Fusion 61 (2021) 066024.

[18] C. Bourdelle, J. F. Artaud, Vbulandi Basiuk, M. Bécoulet, S. Brémond, J. Bucalossi, H. Bufferand, G. Ciraolo, L. Colas, Y. Corre and X. Courtois, WEST physics basis, Nuclear Fusion 55(6) (2015) 063017.

[19] K. Ida, T. Fujita, Internal transport barrier in tokamak and helical plasmas, Plasma Phys. Control. Fusion 60 (2018) 033001.

[20] W. Zhang, P. Jacquet, E. Lerche, R. Bilato, V. Bobkov, D. Coster, Y. Feng, C. Guillemaut, D. Harting, T. Lunt, J.-M. Noterdaeme, G. Szepesi, D. Van Eester, and the JET Contributors, 3D simulations of gas puff effects on edge plasma and ICRF coupling in JET, Nucl. Fusion 57 (2017) 056042.

[21] B. J. D. Tubbing, B. Green, J. How, M. Huart, R. König, C. Lowry, P. Lomas, P. Noll, P. H. Rebut, J. O'Rourke, D. Stork, A. Tanga, A. Taroni, D. Ward, and the JET Team, AC plasma current operation in the JET tokamak, Nucl. Fusion 32 (1992) 967.

[22] M. Romanelli, V. Parail, P. da Silva Aresta Belo, G. Corrigan, L. Garzotti, D. Harting, F. Köchl, E. Militello-Asp, R. Ambrosino, M. Cavinato, A. Kukushkin, A. Loarte, M. Mattei, R. Sartori, Modelling of plasma performance and transient density behaviour in the H-mode access for ITER gas fuelled scenarios, Nucl. Fusion 55 (2015) 093008.

[23] T. M. Wilks, M. Knolker, P.B. Snyder, D. Eldon, F. Scotti, C. Chrystal, F.M. Laggner, C. Lasnier, A. Mclean, T. Osborne, and C. Paz-Soldan, Development of an integrated core–edge scenario using the super H-mode, Nuclear Fusion 61(12) (2021) 126064.

[24] Y. Miao, J. Sheng Hu, C. Fu, Y. Wang, S. A. Khrapak, F. C. Zhong, C. Ran Du, Observations of dust fragmentations in the Experimental Advanced Superconducting Tokamak, J. Plasma Phys. 89 (2023) 175890202.

[25] Y. T. Song, X. L. Zou, X. Z. Gong, A. Becoulet, R. Buttery, P. Bonoli, T. Hoang, R. Maingi, J. P. Qian, X. M. Zhong, A. Liu, E. Li, R. Ding, J. Huang, Q. Zang, H. Liu, L. Wang, L. Zhang, G. Li, Y. Sun, A. Garofalo, T. Osborne, T. Leonard, S. G. Baek, G. Wallace, L. Xu, B. Zhang, S. Wang, Y. Chu, T. Zhang, Y. Duan, H. Lian, X. Zhang, Y. Jin, L. Zeng, B. Lyu, B. Xiao, Y. Huang, Y. Wang, B. Shen, N. Xiang, Y. Wu, J. Wu, X. Wang, B. Ding, M. Li, X. Zhang, C. Qin, W. Xi, J. Zhang, L. Huang, D. Yao, Y. Hu, G. Zuo, Q. Yuan, Z. Zhou, M. Wang, H. Xu, Y. Xie, Z. Wang, J. Chen, G. Xu, J. Hu, K. Lu, F. Liu, X. Wu, B. Wan, J. Li, EAST Team, Realization of thousand-second improved confinement plasma with Super I-mode in Tokamak EAST, Sci. Adv. 9 (2023) 5273.

[26] G. Sinclair, J. K. Tripathi, P. K. Diwakar, A. Hassanein, Melt layer erosion during ELM-like heat loading on molybdenum as an alternative plasma-facing material, Sci. Rep. 7 (2017) 12273.

[27] V. Sizyuk, A. Hassanein, New proposed ITER divertor design using carbon insert on tungsten to mitigate ELMs and secondary radiation effects on nearby components, Sci. Rep. 12 (2022) 4698.

[28] J. Linke, J. Du, Th. Loewenhoff, G. Pintsuk, B. Spilker, I. Steudel, M. Wirtz, Challenges for plasma-facing components in nuclear fusion, Matter Radiat. Extremes 4 (2019) 056201.

[29] C. Bermudo Gamboa, T. Andersson, D. Svensson, F. J. Trujillo Vilches, S. Martín-Béjar, L. Sevilla Hurtado, Modeling of the fracture energy on the finite element simulation in Ti6Al4V alloy machining, Sci. Rep. 11 (2021) 18490.

[30] M. Sun, W. Cao, D. Hu, N. Zhang, R. Chi, Effect of Cover Plate on the Ballistic Performance of Ceramic Armor, Materials 14 (2021) 1.

[31] M. Li, A Fracture Mechanics Study of Tungsten Failure under High Heat Flux Loads, Master's thesis, (2015).

[32] J. Lee, Analysis of Multi-Layered Materials Under High Velocity Impact Using CTH, PhD thesis, p. 198, Feb. (2008).